# Modifications in the frustrated magnetism, oxidation state of Co and magnetoelectric coupling effects induced by a partial replacement of Ca by Gd in the spin chain compound $Ca_3Co_2O_6$

Tathamay Basu, Kiran Singh,* and E.V. Sampathkumaran
Tata Institute of Fundamental Research, Homi Bhabha Road, Colaba, Mumbai 400005, India.

**Abstract**

We have systematically investigated the influence of gradual replacement of Ca by Gd on the magnetic and complex dielectric properties of the well-known geometrically frustrated spin-chain system, $Ca_3Co_2O_6$ ($T_N$= 24 K with additional magnetic transitions below 12K), by studying the series, $Ca_{3-x}Gd_xCo_2O_6$ ($x \leq 0.7$), down to 1.8K. Heat capacity measurements establish that the reduction of $T_N$ with Gd substitution is much less as compared to that by Y substitution. The magnetic moment data reveal that there are changes in the oxidation state of Co as well, unlike for Y-substitution, beyond $x$= 0.2. Thus, despite being isovalent, both these substitutions interestingly differ in changing these magnetic properties in these oxides. We propose that the valence electrons of Y and those of *R* ions play different roles on deciding these magnetic characteristics of these mixed oxides. It is observed that a small amount ($x$= 0.3) of Gd substitution for Ca is enough to suppress glassy *ac* magnetic susceptibility behavior for the peak around 12 K. An additional low-temperature magnetic anomaly close to 5 K gets more prominent with increasing Gd concentration as revealed by heat-capacity data. Trends in temperature dependence of complex dielectric behavior were also tracked with varying composition and a frequency dependence is observed, not only for the transition in the region around 10 K (for some compositions), but also for the 5 K transition which is well-resolved for a higher concentration of Gd. Thus, Gd-substituted $Ca_3Co_2O_6$ series is shown to reveal interesting magnetic and dielectric behavior of this family of oxides.

## 1. Introduction

In the field of geometrical frustration, the spin-chain system $Ca_3Co_2O_6$ [1, 2], crystallizing in the $K_4CdCl_6$-type rhombohedral structure, has generated considerable interest from the angle of fundamental science as well as from the applications potential, for instance, for solid-state fuel-cell and for pigments in coatings [see, for instance, Ref. 3-35, and references therein]. Broadly speaking, this compound undergoes three-dimensional magnetic ordering near ($T_N=$) 24 K with the so-called 'partially-disordered antiferromagnetism (PDAF)' [2], followed by another magnetic transition near 10 K, characterized by a huge frequency dependence of *ac* susceptibility ($\chi_{ac}$). Additional magnetic feature at a lower temperature (around 5-7 K) has also been identified in various measurements [7, 23, 24]. There are several fascinating properties of this compound, the origins of which are under intense discussions in the current literature. Some of these properties are: (i) Complexities in the temperature (*T*) dependence of magnetic structures as well as spin relaxation dynamics [3, 20, 22, 23].; (ii) equally-spaced (about 12 kOe) multiple steps [8, 13, 19, 22, 25, 26] in isothermal magnetization, *M(H)*, where *H* is the externally applied magnetic-field, at very low temperatures (*T* << 10 K), - a rare observation among oxides; (iii) strong magnetoelectric coupling and multiferroicity for this compound [7, 11, 27] and some of its derivatives [29]; (iv) a large orbital moment on Co [28]; and (v) 'incipient spin-chain ordering characteristics near 100 K [6, 12, 14, 23]. All the anomalies arise from the high-spin ($S=2$) trivalent Co ions (CoI) [32] occupying trigonal prismatic site, placed alternately with low-spin ($S=0$) trivalent Co ions (CoII, occupying octahedral site) along *c*-axis. The intrachain (along *c*-axis) interaction is ferromagnetic, whereas interchain interaction (in the basal plane) is antiferromagnetic. The ferromagnetic intrachain interaction (*J*) dominates the antiferromagnetic (*AFM*) interchain interaction (*J'*) [9]. The overall magnetic structure appears to be a complicated one with a modulation along c-axis [3, 4, 17].

The present exhaustive work was primarily undertaken [33] to investigate how the magnetic and magnetoelectric coupling behaviors are influenced by the presence of a strong paramagnetic ion (Gd ions) at the Ca site intervening the Co chains, taking into consideration strong magnetoelectric coupling known for the parent compound [7, 11, 27]. We have earlier reported [34] the influence of non-magnetic substitution, namely, partial Y substitution for Ca, on the magnetic behavior and therefore one of the aims was to compare and contrast the observed magnetic behavior for possible identification of factors responsible for the change of properties, if any, in such chemical substitutional studies on this system. With these motivations, we present the results of a systematic investigation of the series $Ca_{3-x}Gd_xCo_2O_6$ through *ac* and *dc* magnetization, heat-capacity (*C*), and complex dielectric measurements down to 1.8K. Among various observations, the findings of emphasis are: (i) Gd substitution lowers $T_N$ only marginally with respect to that known for Y substitution [34]. There is a change in the magnetic moment behavior as well, as though the valence of CoI changes with Gd substitution; interestingly, this seems to change even for CoII site in the case of *x*= 0.7. These findings are broadly different from those observed for Y-substituted systems, whereas, in general, such isovalent rare-earth substitutions are expected to bring out similar effect on the magnetism of parent compounds. In that sense, the present results are interesting. We propose that the changes induced by valence electrons of Y and other rare-earth ions are different on these magnetic properties in this series. (ii) A small doping of Gd for Ca site (i.e., x>0.2) is sufficiemt to suppress spin-glass-like magnetic behaviour (as inferred from $\chi_{ac}$ data) near 12 K in favour of more ordered magnetism; it appears that the onset of magnetic ordering could be described in terms of antiferromagnetism in these compositions, whereas a lower temperature transition (at ~ 5K) is dominated by a ferromagnetic interaction, as revealed by the influence of magnetic-field on the peak-temperature in the *C(T)* curves. (iii) With respect to complex dielectric property, the features in *M* track those in dielectric properties indicating the existence of magnetoelectric coupling in the entire composition range; in particular, dielectric constant ($\varepsilon'$) data reveal distinct frequency dependence of a shoulder near 5K for higher dopings, apart from that known for the parent compound in the vicinity of transition near 10 to 12K, as though relaxor ferroelectric-like behavior is developed for all transitions well below $T_N$.

## 2. Experimental details

Polycrystalline samples of the series $Ca_{3-x}Gd_xCo_2O_6$ (*x*= 0.0, 0.1, 0.2, 0.3, 0.5, 0.7 and 1.0) were prepared by solid state reaction method as described earlier [25] starting with stoichiometric amounts of respective high-purity (>99.95%) oxides, $CaCO_3$, $Gd_2O_3$ and $Co_3O_4$. The samples were characterized by x-ray diffraction (XRD, Cu $K_\alpha$ ) and Rietveld refinement established that all the compositions except *x*= 1.0 form in a single phase with proper structure and therefore the investigations are restricted to *x* ≤ 0.7. There is a marginal decrease of lattice parameter '*a*' with increasing *x*, but '*c*' increases (see Table 1). An expansion of the unit-cell with Gd substitution is obvious from a visual inspection of diffraction peak positions as well (see, for example, figure 1).

*Dc* magnetization measurements as a function of *T* (1.8-300 K) were carried out in a magnetic field of 5 kOe with the help of a commercial superconducting quantum interference device (SQUID, Quantum Design) magnetometer. The $\chi_{ac}$ was measured with selected frequencies (*v*=1.3, 13, 133 and 1339Hz) with the Quantum Design magnetometer. Isothermal *M(H)* curves at different temperatures were obtained using a vibrating sample magnetometer (VSM, Quantum Design) in a range of *H* = 0-140 kOe. The heat-capacity measurements were performed as a function of temperature in different magnetic fields (up to 140 kOe) employing a Physical Property Measurements System (PPMS, Quantum Design). The frequency (1 - 100 kHz) dependence of complex dielectric permittivity was carried out using E4980A LCR meter (Agilent Technologies) with a home-made sample-holder which is coupled to this PPMS.

## 3. Results

### 3.1 Heat capacity

Fig.2 shows *C/T* as a function of *T* measured in the presence of various magnetic fields. Focussing on the zero-field data, for the parent compound, with decreasing temperature, there is an upturn below about 24 K followed by a peak at the onset temperature of PDAF. There is no other well-defined peak at lower temperatures [23, 35]. With increasing doping of Gd, this peak shifts to lower temperatures marginally to about 21 and 17 K for *x*= 0.1 and 0.2 respectively. With further higher doping, for *x*= 0.3 to 0.7, this peak gets broadened and remains around 13-15 K. Thus, one can infer that the onset of magnetic ordering is decreased to a marginally lower temperature with Gd doping. However, in contrast to the behavior of the parent compound, an additional weak upturn below about 7 K followed by a broad peak around 5 K appears with Gd doping. This indicates that Gd doping sharpens the features due to additional magnetic transitions occurring at very low temperatures (well below $T_N$), as though spin-glass-like description becomes gradually less relevant for this transition.

In order to throw more light on the nature of these transitions, we obtained *C(T)* in the presence of magnetic fields. For the parent component, it has been known that there is a finite (about 0.5 K) upward shift of the peak temperature for *H*= 30 kOe [35]. This upward shift is unexpected for antiferromagnetism. Therefore this finding signals dominant role of intrachain ferromagnetic correlations at the onset of magnetic ordering in this compound, despite the fact that the net magnetic structure is antiferromagnetic. This feature is found to persist for *x*= 0.1 only (which could be seen if the peak positions are expanded for 30 and 50 kOe). However, for $x \geq 0.2$, this temperature is found to undergo a marginal suppression, as seen in figure 2, with increasing *H*. We interpret that this is due to increasing dominance of interchain antiferromagnetic correlations. (Also, with respect to the magnitude at the peak temperature, in the parent compound, this first increases and then decreases with increasing magnetic field, the reason for which lies in subtle differences in the behavior of magnetic entropy. However, this anomaly also vanishes for $x \geq 0.3$). With respect to the feature below 7 K, the peak is found to shift towards higher temperatures for all Gd substituted specimens. These results suggest that the lower transition is not dominated by antiferromagnetic correlations, but consistent with a stronger influence from ferromagnetic correlations. Clearly, the magnetic origin of these two features is of different type. An inspection of the curves for *x*= 0.5 and 0.7 for 80 kOe clearly reveals that the features due to both the transitions tend to merge.

### 3.2. Dc magnetization

Temperature dependencies of *dc* magnetic susceptibility ($\chi_{dc}$) for the field-cooled (*FC*) and zero-field-cooled (*ZFC*) conditions of the specimens at low temperatures are shown in figure 3. As known in the literature for an application of 5 kOe [25], for *x*= 0, there are distinct changes in the slopes with decreasing temperature: An upturn around ($T_N$=) 24 K, a peak near 12 K and a sharper fall near 4-7 K in the ZFC curve due to changes in magnetic character, with a bifurcation of ZFC-FC curves near 12 K where spin-glass-like anomaly has been known to set in [24, 25]. For *x*= 0.1, there is a clear evidence for the depression of $T_N$ marginally in this data as well. However, with a further increase of Gd concentration, the sharp upturn at $T_N$ is smoothened and one sees only a more gradual change in the slope of the curve. Therefore, it is difficult to infer the trend in $T_N$ from the $\chi_{dc}$ data. Though other features like the peaks in ZFC curves and the bifurcation of ZFC-FC curves persist on Gd substitution, there are qualitative changes in the shapes of the curves. It is rather difficult to infer any further meaningful conclusions based on these curves and therefore these results are essentially presented for the sake of completeness.

With respect to the data in the paramagnetic state (not shown in the form of a figure), there is a deviation from high temperature linear behavior below 140 K due to 'incipient' spin-chain ordering [6, 12]. The sign of the paramagnetic Curie temperature ($\Theta_P$), if derived restricting the Curie-Weiss fit to the region 160-300 K, is found to be positive for all compositions. However, the values are found to be gradually reducing with increasing x (see Table 1). The trend in the values of $\Theta_P$ (that is, a gradual reduction) indicates decreasing strength of intrachain ferromagnetic correlation with increasing *x*. The values of effective magnetic moment ($\mu_{eff}$) are also given in Table 1. Assuming theoretical paramagnetic moment of Gd ($7.94\mu_B$), the deduced values of $\mu_{eff}$ on Co are listed in Table 1. The value of $\mu_{eff}$ per Co formula unit is found to remain constant till *x*= 0.2, subsequently undergoing a notable reduction with a further Gd substitution (assuming charge-balance, even one accounts for a small change in oxygen content due to the substitution of a trivalent ion for a divalent ion) for *x*= 0.3 and 0.5. We attribute this decrease in the value of $\mu_{eff}$ on Co to the transfer of Gd valence electron to the CoI ion occupying trigonal prismatic site (that is, a change of high-spin $d^6$ to high-spin $d^7$) for these two compositions. It is however interesting to note that the value increases for $x= 0.7$ ($\mu_{eff} \sim 4.87$ $\mu_B$). We interpret this in terms of a change in electronic configuration at the octahedral site as well, thereby making it (low-spin) $d^7$ configuration (which is characterized by a non-zero magnetic moment) at this site for this composition. Thus, it appears that there is a change in the valence state of Co following Gd substitution beyond $x= 0.2$.

We have obtained more convincing evidence for the change in the oxidation state of Co from the isothermal *M(H)* data in the magnetically ordered state. We show the data in Fig. 4 for 1.8 and 8 K for selected Gd-substituted compositions. Needless to elaborate that the multiple steps present in the single crystals of the parent compound at 1.8K are known to be smeared in the polycrystalline form, which can be further smeared by any chemical substitution. Therefore, a reasonable approach to look for such steps for the Gd substituted compositions in this study is to take the derivative plots of the experimental data. Thus, we inferred the existence of such steps for $x= 0.1$. This is not shown in the form of a figure, as this is not a point of interest in this work. The plots are found to be hysteretic, but the area within the hysteresis curve decreases with increasing $T$ and/or $x$. We now return to the point of emphasis in the *M(H)* data. In order to infer whether the magnetic moment on Co is influenced in the magnetically ordered state, in the same figures, we have shown the *M(H)* plots, derived by summing experimentally observed *M(H)* of $Ca_3Co_2O_6$ and theoretically calculated *M(H)* for paramagnetic Gd ion using Brillouin function. It is found that the experimental high-field (120 kOe) values of magnetization ($M_h$) match reasonably well with the calculated ones for x=0.1 (and also for x=0.2, not shown) at all temperatures within the experimental error, but $M_h$ values are found to be reduced when the concentration of Gd is increased. This means a significant reduction in the magnetic moment on Co for $x= 0.3$ and 0.5, which is consistent with the addition of an electron to the CoI site (that is, some ions at the CoI site are in S=3/2 state because of $d^7$ high-spin configuration), supporting the inference from the paramagnetic susceptibility data. The $M_h$ value for the $x$=0.7 composition is also further decreased with respect to the derived value, but the magnitude of the decrease is much larger compared to that for other compositions. If the electron added to CoII couples parallel to the spins of CoI, one would have expected an increase of $M_h$ with respect to that for $x= 0.3$. Therefore, we conclude that the magnetic moments at CoI and CoII sites couple antiparallel to each other for this composition [25].

### 3.3. Ac susceptibility

The parent compound has been known [24, 25] to show a broad peak in *ac χ* (Fig. 5a, 5b) in the vicinity of the 10-12K transition with a strong *ν*-dependence, which makes the observed glassy behavior different from that in conventional spin-glasses [36]. For instance, the peak temperature in the real part ($\chi'$) moves from 10-12 K for $\nu= 1.3$ Hz to about 17 K for $\nu= 1.339$ kHz, whereas for conventional spin-glasses, corresponding variation is usually much smaller than 1 K. (Possible features at $T_N$ could not be resolved; however a gradual change in slope could be observed around this temperature [25]). Clearly, spin-dynamics of this compound in the low temperature range is fascinating as mentioned at the Introduction. We have investigated how this *ac χ* behavior gets modified with Gd substitution. These results are shown in figure 5. The features mentioned above are observable for $x= 0.1$ and 0.2, in both real and imaginary ($\chi''$) parts (see figures 5c, 5d, 5e and 5f); however, the intensity of the peak comes down gradually with increasing $x$; one possible explanation for this is that these compositions are magnetically inhomogeneous with the parent-compound-like behavior being depressed for Co ions surrounded by Gd. With a further increase of Gd concentration, the $\chi''$ signal completely vanishes (see figures 5h, 5j, and 5l), though the signal for the real part is still observable with a peak near 13 K. Absence of $\chi''$ signal is a sufficient condition [36] to conclude that the magnetic ordering in this temperature range can not arise from spin-glass freezing. Therefore, the observation of finite signal in the real part alone implies that there is a non-glassy magnetic ordering for $x>0.2$. This finding is consistent with the observations from the *C(T)* data for such a low-*T* range.

Another observation we have made from a careful look at the data in figure 5g for $x= 0.3$ is that an additional shoulder appears near 7 K below the peak temperature (in $\chi'$). It is of interest to note that the signal intensity decreases with increasing frequency for this transition (for higher *x*) and this could be an artefact of a small frequency dependence, with this peak moving towards high temperatures with increasing frequency . This shoulder is prominently seen for x= 0.7 also (figure 5k), which is consistent with heat capacity data on the existence of another transition around this temperature.

### 3.4. Dielectric permittivity

In figure 6, we show the frequency dependence of dielectric constant as well as loss factor (*tan$\delta$*) as a function of temperature in the absence of any externally applied magnetic field. It appears that the parent compound exhibits multiglass-like behavior [37-40]. We plan to discuss the behavior of the parent compound in detail in a future publication. Therefore, here we reproduce only those data and discussions which are crucial for the comparisons with the behavior in Gd-substituted compounds. The parent compound shows a huge frequency dependent behaviour around 17-29K, in both $\varepsilon'$ and *tan$\delta$*, (see Fig. 6a and 6b). There is a minimum in the intermediate *T*-range followed by an upturn beyond 100 K in these parameters, which can be attributable to a small change in electrical resistivity in this temperature range, though this compound is still highly insulating as characterized by low values of *tan$\delta$* as well. It appears that the resolution of the features before and after the onset of magnetic ordering depends sensitively on the sample, as this resistivity behavior itself could be sample-dependent due to varying levels of defects and grain-boundaries influencing hopping conductivity [41]. For instance, there are clearly two peaks below $T_N$ as reported in Ref. 27, whereas we could infer these two anomalies below 24 K from the derivative curve only (see Fig. 6c, also for a distinct anomaly at $T_N$ ). However, the minimum mentioned above is not resolved in Ref. 27. For *x*= 0.1 (Fig. 6d), the above features are modified dramatically due to possible tendency to less insulating behavior, as indicated by a weak increase in *tan$\delta$* as well, even beyond 25 K (figure 6e), unlike in the parent compound. As a result, the 17-20 K peak manifests itself in the form of a shoulder in the curves in this *T*-range. Interestingly enough, for a higher doping, there is a further qualitative change in the shapes of the curves. For instance, for *x*= 0.3 (Fig. 6f), in the $\varepsilon'(T)$ curves, the feature around 20 K is unresolvable (presumably manifesting as a change in slope only), but a sudden drop appears at lower temperatures, as though there is a transition near 5 K (as revealed by the peak in *tan$\delta$* for the lowest frequency, figure 6g). The frequency dependence however is relatively weak (about 3 K for 1 to 100 kHz variation). Since 20K-feature is weak, the peak due to this 5K-transition is well-resolved in *tan$\delta$* curves. Similar features are seen for *x*= 0.5 (not shown) as well. Thus, we essentially observe only one (weakly) frequency dependent component (which is near 5 K), unlike for the parent case. Interestingly enough, for *x*= 0.7 (Fig. 6h), the 17-20 K feature reappears, in addition to the one below 7 K featuring as a shoulder, consistent with that observed in *C(T)* data. We tend to believe that this resolution of peaks is a clear manifestation of the change in electrical conductivity with varying composition with a change in the oxidation state of Co at two different sites (as described earlier) due to Gd substitution. Thus, the dielectric behavior appears to be more sensitive to any subtle change in electronic configuration in this system.

We have also measured ε′ and *tanδ* as a function of applied external magnetic fields up to 140 kOe for all compositions. We find a broad correspondence between *M(H)* and these data, as demonstrated for the parent compound in the literature [7, 27]. For the sake of brevity, we show the data at 2 K in figure 7, for instance, for *x*= 0.1, 0.5 and 0.7, for which we argued above that there is a change in the magnetic character of Co across these compositions. For x= 0.1, as in the parent compound, there are distinct changes in slopes of *M(H)* near 12, 24, 36 and 40-50 kOe (for increasing magnetic fields, Fig. 7g)). Corresponding changes of slopes are seen in the derivative curves of ε′(*H*) also (Fig. 7j). There is a change in slope in *M(H)* in the vicinity of 25 - 40 kOe in all these data for *x*= 0.5 and 0.7 as well (Fig 7h, 7k; 7i, 7l). The vertical dashed lines are shown for derivative curves in figure 7 to highlight this point. Hysteretic effects are also reflected in both these measurements as in magnetization curves. These clearly establish the existence of magnetelectric coupling in the entire solid solution.

## 4. Discussion

An important observation, that is distinctly apparent from the *C(T)* data, is that there is a gradual depression in the magnetic ordering onset temperature with increasing Gd concentration, reaching a value around 15 K for *x*= 0.7. In sharp contrast to this, isovalent *Y* substitution for Ca brings out [34] a much sharper depression of $T_N$, for instance, close to 4 K for *x*= 0.7. This is despite the fact that the values of ionic radii of both Y and Gd are very close, which is reflected in trends in the lattice constants as well. That is, *c*-axis gets elongated significantly and *a*-axis shrinks relatively weakly, for both the substitutions. Another distinct difference between both the substitutions is that there is a relatively large decrease in the magnetic moment per formula unit as *x* is changed to 0.5 from 0.0 for the Gd series; such a decrease has been reported on the basis of neutron diffraction studies on $Ca_{2.75}R_{0.25}Co_2O_6$ (*R*= Dy, Lu) [21]. However, corresponding change was found to be negligible for the Y series. All these findings suggest that the electronic structure changes induced by *R* ion substitution for Ca is different from that induced by Y substitution in this series, thereby bringing out differences in net magnetism. This situation is somewhat different from that normally encountered for such Y and other rare-earth ion substitutions in magnetic materials, in which case such trivalent ionic substitutions are known to bring out similar changes An interesting observation is that there is a non-monotonic increase of the effective moment, as seen for *x*= 0.7 in the Gd series, which was also not observed for the Y series. Clearly, as pointed out earlier, Gd substitution decreases the oxidation state at the CoI site for *x*= 0.5, followed by a decrease of the oxidation state at the CoII site for *x*= 0.7.

With respect to the magnetic features at lower temperatures (<< $T_N$), a comparison of the figure 5 in Ref. 34 and figure 5 in this article reveals that the behavior of $\chi_{ac}$ are apparently similar in both cases. That is, there is a dramatic suppression of the intensity of the 12 K feature with an increase in *x*. However, a feature (possibly with a weak frequency dependence) at further lower temperatures for the Gd-based series is clearly resolved. This could endorse some claims in the literature [7, 23, 24] that, in addition to 10-12K transition, there is an additional transition at a lower temperature even for the parent compound; this transition in ac χ is possibly masked for the parent compound by the relatively intense signal from the 10-12 K transition. This could be the reason why the ac χ peak for the parent compound is quite broad, as it appear to be a superimposition of at least two closely-spaced peaks. From the suppression of χ" for higher Gd content, we infer that these lower temperature transition are of a non-spin-glass-type. In short, these results reveal suppression of low-temperature complex spin-glass-like anomalies towards a non-glassy magnetic ordering with increasing Gd content. This conclusion is endorsed by a sharpening of the *C/T* feature in this temperature region with increasing *x*; the fact that this peak in *C/T* gets shifted to a higher temperature with increasing magnetic field indicates increasing dominance of a ferromagnetic component.

With respect to complex dielectric behavior, it is sufficiently established that magnetoelectric coupling is intrinsic in its character for the parent compound [7, 11, 27] and this work presents enough evidence for its persistence in the entire solid solution. However, a noteworthy conclusion derived from the two peaks in the *T*-dependent data was that there are two relaxation times. We can not extend this conclusion for most of the compositions in the present solid solution as two peaks in the dielectric data could not be resolved and also there is a change in the oxidation state of Co at the trigonal prismatic site. However, for *x*= 0.7, interestingly, a strong frequency dependence of $\varepsilon'$ is noted even around 15 K for *x*= 0.7 (without such a frequency dependent behavior in ac χ). This is in addition to a weak frequency dependent feature below 10 K (as in ac χ) seen for all Gd concentrations. Clearly, relaxation dynamics gets more complex with Gd substitution. We hope this observation offers a theoretical challenge to understand how such multiple ν-dependent complex dielectric features appear for *x*= 0.7 despite a change in the oxidation state of Co ions. Another finding of interest is that, as seen in figure 7, the derivative curves $d\varepsilon'/dH$ show maxima and minima, which could imply the existence of a competition between antiferromagnetic correlations and ferromagnetic correlations with varying magnetic-field in controlling dielectric behavior, at least for this family of materials as argued by Kaushik et al [29]. Finally, it may be recalled that another derivative of $Ca_3Co_2O_6$, viz., $Ca_3CoMnO_6$, has been reported by Choi et al [29] to undergo multiferroic behavior through exchange striction associated with up-up-down-down magnetic order of $Co^{2+}$ and $Mn^{4+}$. In contrast to this, the present compound is characterized by a complex ultraslow evolution between two long-range magnetically ordered states [17], which apparently gets modified with the substitution of Gd for Ca due to changes in the oxidation state of Co as discussed in this article. It is therefore of great interest to perform deeper investigations of multiferroicity in this solid solution.

5. **Conclusion**

We have systematically studied the influence of a disruption of the surrounding of spin-chains on the geometrically frustrated magnetism and magnetoelectric coupling property of $Ca_3Co_2O_6$, by a gradual replacement of Ca by a paramagnetic lanthanide ion, namely, in the series, $Ca_{3-x}Gd_xCo_2O_6$ ($x \leq 0.7$). The results enable us to infer the presence of at least three characteristic temperatures (24 K, 10-12 K and 5 K) around which anomalies in different physical properties exist. The results reveal that the three dimensional magnetic ordering temperature (which is close to 24 K for the parent compound) gets suppressed with Gd substitution, but the degree of suppression is relatively less when compared to that brought out by isoelectronic Y substitution. In addition, with increasing Gd concentration, the oxidation state of Co at the trigonal prismatic site is initially decreased, while the one at the octahedral site also appears to be affected for a higher concentration of Gd. These changes, which are different from those brought out by Y substitution despite having the same oxidation state as Gd ion, - somewhat different from what one usually encounters in magnetic materials - imply that the roles of valence orbitals of lanthanide ions and Y on the magnetic moment of Co and the onset of magnetic ordering are different in this system. However, surprisingly, spin-glass-like anomalies around 10-12 K observed for the parent compound gets suppressed with Gd substitution similar to that observed for Y substitution, with a ferromagnetic-like component developing at lower temperatures. We observe in general magnetolectric coupling in the entire solid solution and it appears that ferrolectric-relaxor-like behavior is observed at low temperatures. Thus, this solid solution exhibits interesting magnetic and magnetoelectric behavior.

Acknowledgment:

The authors thank Kartik K Iyer for his help during measurements.

**References**

*Present address: UGC-DAE Consortium for Scientific Research, University Campus, Khandwa Road, Indore - 452017, India.

1. S. Aasland, H. Fjellvag, and B. Hauback, Solid State Commun. **101** (1997) 187.

2. H. Kageyama, K. Yoshimura, K. Kosuge, H. Mitamura, and T. Goto, J. Phys. Soc. Jpn. **66** (1997) 1607.

3. S. Agrestini, C. Mazzoli, A. Bombardi, and M.R. Lees, Phys. Rev. **B 77** (2008) 140403(R).

4. S. Agrestini, L.C. Chapon, A. Daoud-Aladine, J. Schefer, A. Gukasov, C. Mazzoli, M.R. Lees, and O.A. Petrenko, Phys. Rev. Lett. **101** (2008) 097207.

5. A. Bombardi, C. Mazzoli, S. Agrestini, and M. R. Lees, Phys. Rev. B **78** (2008) 100406(R).

6. P.L. Paulose, N.Mohapatra, and E.V. Sampathkumaran, Phys. Rev. B **77** (2008) 172403.

7. N. Bellido, C. Simon, and A. Maignan, Phys. Rev. B **77** (2008) 054430.

8. Yu. B. Kudasov et al, Phys. Rev. B 78, 132407 (2008).

9. L. C. Chapon, Phys. Rev. B **80**, 172405 (2009).

10. R. Soto, G. Martinez, M.N. Baibich, J.M. Florez, and P. Vargas, Phys. Rev. **B 79,** 184422 (2009).

11. N. Bellido, C. Simon, and A. Maignan, J. Magn. Magn. Mater. **321,** 1770 (2009)**.**

12. M. R. Bindu, K. Maiti, S. Khalid, and E.V. Sampathkumaran, Phys. Rev. **B 79,** 094103 (2009); Smita Gohil, Kartik K Iyer, P. Aswathi, Shankar Ghosh, and E.V. Sampathkumaran, J. App. Phys. **108,** 103517 (2010).

13. M. H. Qin, K.F. Wang, and J.M. Liu, Phys. Rev. B **79,** 172405 (2009).

14. Niharika Mohapatra, Kartik K Iyer, Sitikantha D Das, B.A. Chalke, S.C. Purandare, and E.V. Sampathkumaran, Phys. Rev. B **79,** 140409(R) (2009).

15. J.-G. Cheng, J.-S. Zhou, and J. B. Goodenough, Phys. Rev. B **79,** 184414 (2009).

16. Y. Shimizu, M. Horibe, H. Nanba, T. Takami, and M. Itoh, Phys. Rev. B **82**, 094430 (2010).

17. S. Agrestini, C. L. Fleck, L. C. Chapon, C. Mazzoli, A. Bombardi, M. R. Lees, and O. A. Petrenko, Phys. Rev. Lett. **106**, 197204 (2011).

18. G. Allodi, R. De Renzi, S. Agrestini, C. Mazzoli and M.R. Lees, Phys. Rev. B, **83**, 104408 (2011).

19. Yu. B. Kudasov, A.S. Korshunov, V.N. Pavlov, and D.A. Maslov, Phys. Rev. B **83**, 092404 (2011).

20. Y. Kamiya and C.D. Batista, Phys. Rev. Lett. **109,** 067204 (2012).

21. Anil Jain, S. M. Yusuf, S. S. Meena, and Clemens Ritter, Phys. Rev. B **87**, 094411 (2013).

22. V. Hardy, D. Flahaut, M. R. Lees, and O. A. Petrenko, Phys. Rev. B **70**, 214439 (2004).

23. V. Hardy, S. Lambert, M. R. Lees, D. McK. Paul, Phys. Rev. B **68**, 014424 (2003)

24. A. Maignan, C. Michel, A.C. Masset, C. Martin, and B. Raveau, Eur. Phys. J. **B 15** (2000) 657.

25. S. Rayaprol, K. Sengupta. And E.V. Sampathkumaran, Solid State Commun. **128,** 79 (2003).

26. V. Hardy, M.R. Lees, O.A. Petrenko, D. McK Paul, D. Flahaut, S. Hebert, and A. Maignan, Phys. Rev. B **70**, 064424 (2004); A. Maignan, V. Hardy, S. Hebert, M. Drillon, M.R. Lees, ). Petrenko, D. McPaul, and D. Khomskii, J. Mater. Chem. **14**, 1231 (2004).

27. P.L. Li, X.Y. Yao, K.F. Wang, C.L. Lu, F. Gao, and J.M. Liu, J. App. Phys. **104**, 054111 (2008).

28. T. Burnus, Z. Hu, M.W. Haverkort, J.C. Cezar, D. Flahaut, V. Hardy, A. Maignan, N.B. Brookes, A. Tanaka, H.H. Hsieh, H.-J. Lin, C.T. Chen, and L.H. Tjeng, Phys. Rev. B **74,** 245111 (2006); H. Wu, M.W. Haverkort, Z. Hu, D.I. Khomskii, and L.H. Tjeng, Phys. Rev. B **95,** 186401 (2005).

29. Y.J. Choi, H.T. Yi, S. Lee, Q. Huang, V. Kiryukhin, and S.-W. Cheong, Phys. Rev. Lett. **100,** 047601 (2008); V. Kiryukhin, S. Lee, W. Ratcliff, Q. Huang, H.T. Yi, Y.J. Choi, and S.-W. Cheong, Phys. Rev. Lett. **102,** 187202 (2009); S. D. Kaushik, S. Rayaprol, J. Saha, N. Mohapatra, V. Siruguri, P.D. Babu, S. Patnaik, and E.V. Sampathkumaran, J. App. Phys. **108**, 084106 (2010).

30. T. Wei, Sci. Rep. **3,** 1125 DOI: 10.1038/s rep 01125.

31. O.A. Fouad, J. Alloys and Comp. **537,** 165 (2012).

32. E. V. Sampathkumaran, N. Fujiwara, S. Rayaprol, P. K. Madhu and Y. Uwatoko, Phys. Rev. B **70**, 014437 (2004).

33. Priliminary results have been reported at the57th DAE Solid State Physics symposium 2012, Tathamay Basu, Kartik K Iyer, P.L. Paulose, and E.V. Sampathkumaran, AIP Conf. Proc.1512, 1150 (2013).

34. S. Rayaprol and E.V. Sampathkumaran, Pramana – J. Phys. **65**, 491(2005).
35. E.V. Sampathkumaran, Z. Hiroi, S. Rayaprol, and Y. Uwatoko, J. Magn. Magn. Mater. **284,** L7 (2004).
36. K. Binder and A.P. Young, Rev. Mod. Phys. **58,** 801 (1986).

37. See, for example, Anar Singh, Vibhav Pandey, R.K. Kotnala, and Dhananjay Pandey, Phys. Rev. Lett. **101**, 247602 (2008).

38. V.V. Shvartsmnn, S. Bedanta, P. Borisov, W. Kleeman, A. Tkach, and P.M. Vilarinho, Phys. Rev. Lett. **101,** 165704 (2008);

39. D. D. Choudhury, P. Mandal, R. Mathieu, A. Hazarika, S. Rajan, A. Sundaresan, U.V. Waghmare, R. Knut, O. Karis, P. Nordblad, and D. D. Sarma, Phys. Rev. Lett. **108,** 127201 (2012); K. Singh, A. Maignan, C. Simon, S. Kumar, C. Martin, O. Lebedev, S. Turner, and G. Van Tendeloo, J. Phys.: Condens. Matter. **24**, 226002 (2012).

40. Y. Yamaguchi, T. Nakano, Y. Nozue, and T. Kimura, Phys. Rev. Lett. **108,** 057203 (2012).

41. B. Raquet, M.N. Baibich, J.M. Broto, H. Rakoto, S. Lambert, and A. Maignan, Phys. Rev. B **65**, 104442 (2002).

Table 1: The lattice constants, *a* and *c*, paramagnetic Curie temperature ($\Theta_P$) and effective moment per formula unit obtained from the Curie-Weiss in the region 160 - 300 K, and the magnetic moment on Co in the series, $Ca_{3-x}Gd_xCo_2O_6$.

| x | a (Å) (± 0.004 Å) | c (Å) (± 0.004 Å) | $\Theta_P$ (K) ± 1 K | $\mu_{eff}$ (± 0.1$\mu_B$/f.u) | $\mu_{eff}$ on Co ion ($\mu_B$) |
|---|---|---|---|---|---|
| 0.0 | 9.082 | 10.387 | 34 | 5.1 | 5.1 |
| 0.1 | 9.078 | 10.417 | 30 | 5.6 | 5.0 |
| 0.2 | 9.071 | 10.444 | 23 | 6.1 | 4.97 |
| 0.3 | 9.076 | 10.515 | 12 | 6.3 | 4.54 |
| 0.5 | 9.067 | 10.555 | 9 | 7.2 | 4.51 |
| 0.7 | 9.051 | 10.593 | 2 | 8.2 | 4.87 |

# Figure captions

**Fig.1**: X-ray diffraction pattern (Cu $K_\alpha$) for the series, $Ca_{3-x}Gd_xCo_2O_6$. The pattern for one plane is shown in an expanded form to show expansion of unit-cell with increasing $x$. The asterisk in the pattern for $x= 1.0$ is an extra line.

**Fig. 2:** Heat-capacity divided by temperature as a function of temperature for the series, $Ca_{3-x}Gd_xCo_2O_6$.

**Fig.3:** *Dc* magnetization as a function of temperature measured in a magnetic field of 5 kOe for the series, $Ca_{3-x}Gd_xCo_2O_6$, for zero-field-cooled and field-cooled conditions of the specimen. The lines through the data points serve as guides.

**Fig.4:** Isothermal magnetization curve as a function of magnetic field at various temperatures for (a) x= 0.1, (b) x= 0.3, (c) 0.5 and (d) 0.7 in the series, $Ca_{3-x}Gd_xCo_2O_6$, at T=1.8 and 5 K. The continuous curves are obtained experimentally and dashed lines are derived, as discussed in text. The arrows are marked to show the direction of change of magnetic-field.

**Fig. 5.** Real ($\chi'$) and imaginary ($\chi''$) parts of ac susceptibility for the series, $Ca_{3-x}Gd_xCo_2O_6$. The line through the data points serve as guides to the eyes.

**Fig. 6:** Dielectric constant and *tanδ* as a function of temperature at selected frequencies for selected compositions in the series, $Ca_{3-x}Gd_xCo_2O_6$. (a, b) $x= 0.0$, (d, e) $x= 0.1$, (f, g) 0.3, and (h, i) $x= 0.7$. The derivative curve for $x= 0$ for one frequency (50 kHz) is also shown in (c).

**Fig. 7:** Isothermal dielectric constant and loss factor for the compositions $x= 0.1$. 0.5 and 0.7 of the series, $Ca_{3-x}Gd_xCo_2O_6$ at 2K. Derivative of isothermal magnetization and dielectric constants (for increasing field-direction only) are also shown; dashed vertical line is drawn to show the magnetic field around which metamagnetic anomalies are seen; for the sake of clarity, the lines through the data points only are shown for the sake of clarity and hence look noisy.

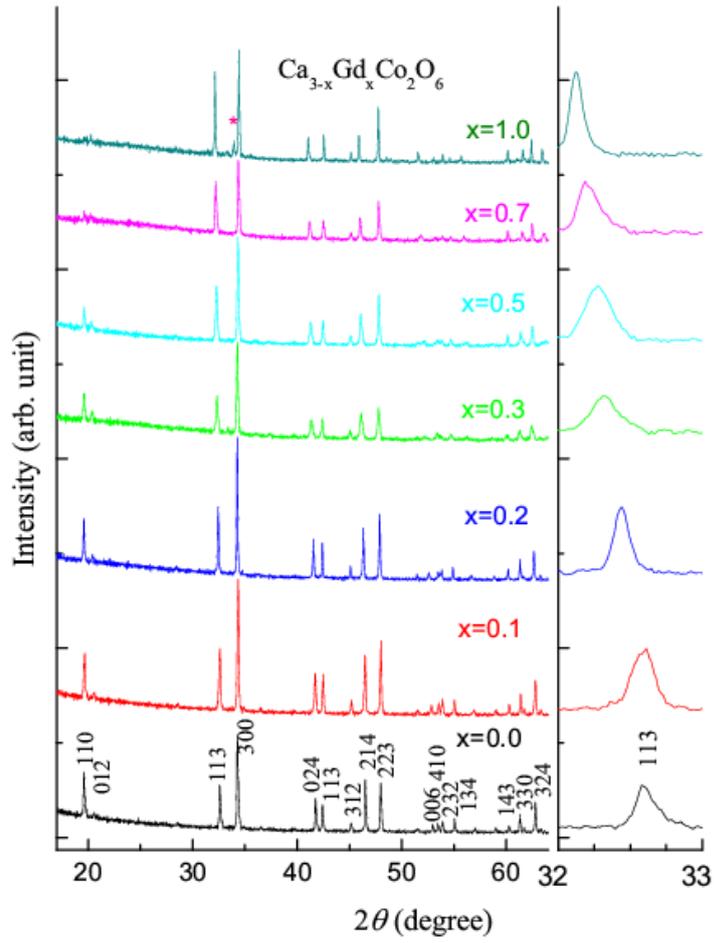

Figure 1

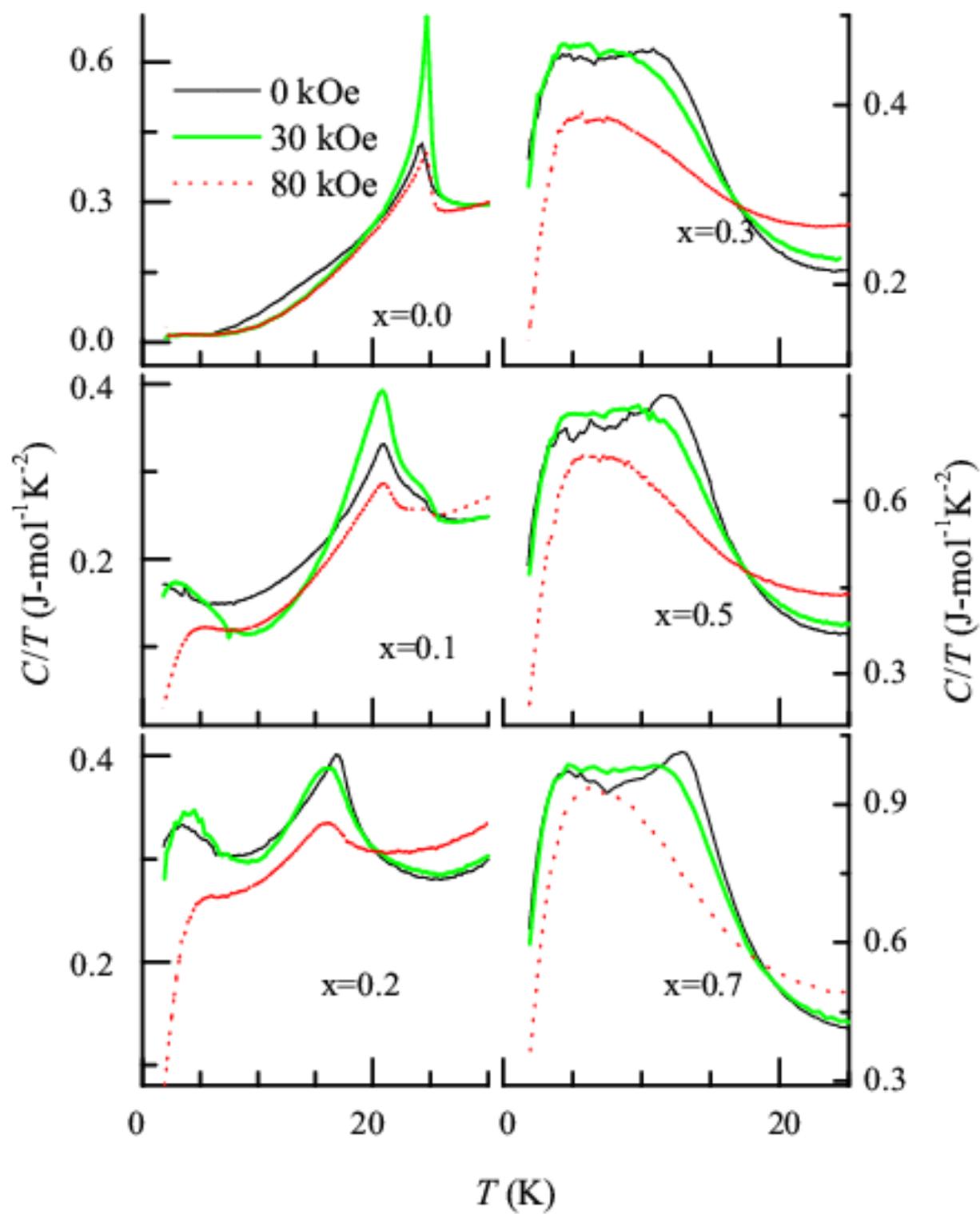

Figure 2

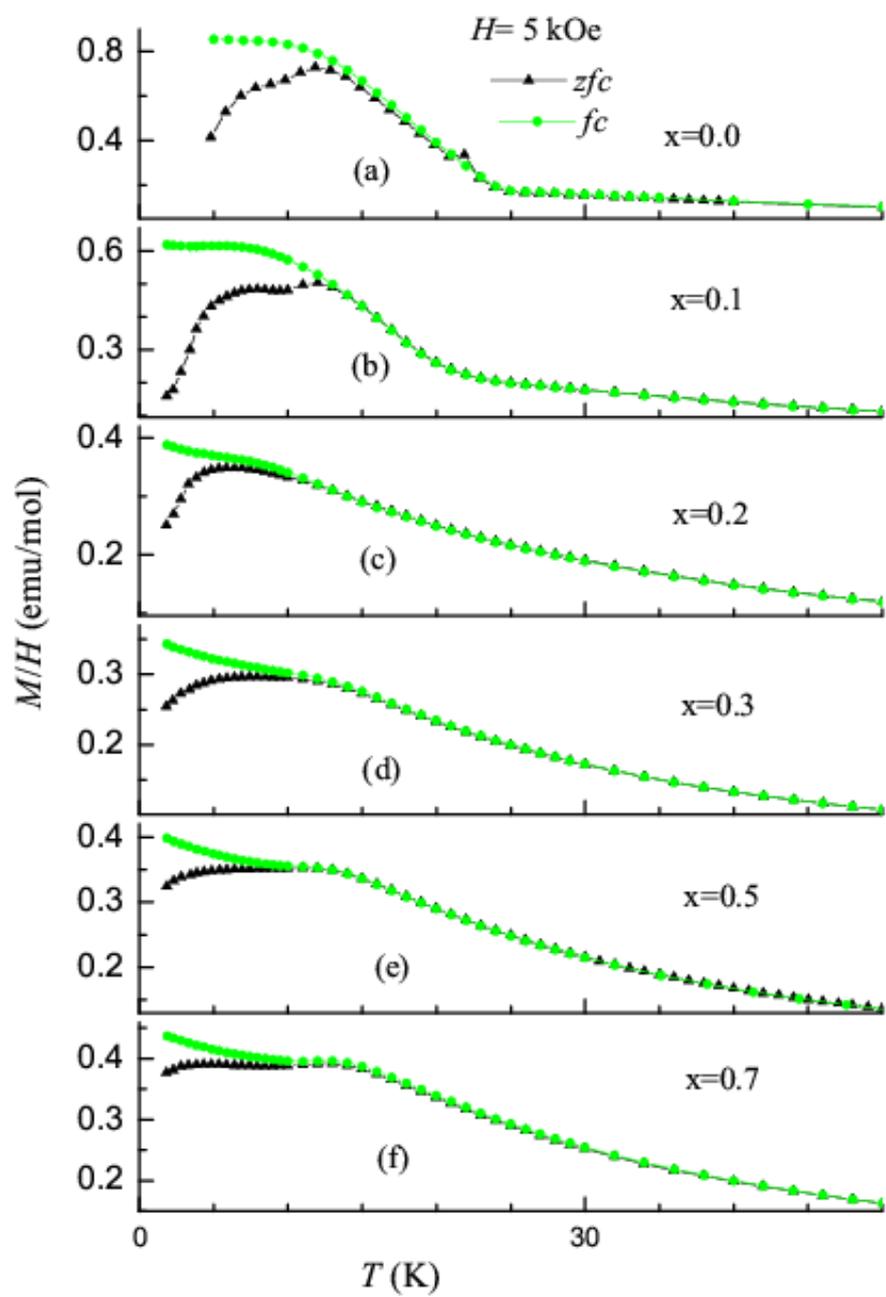

Figure 3

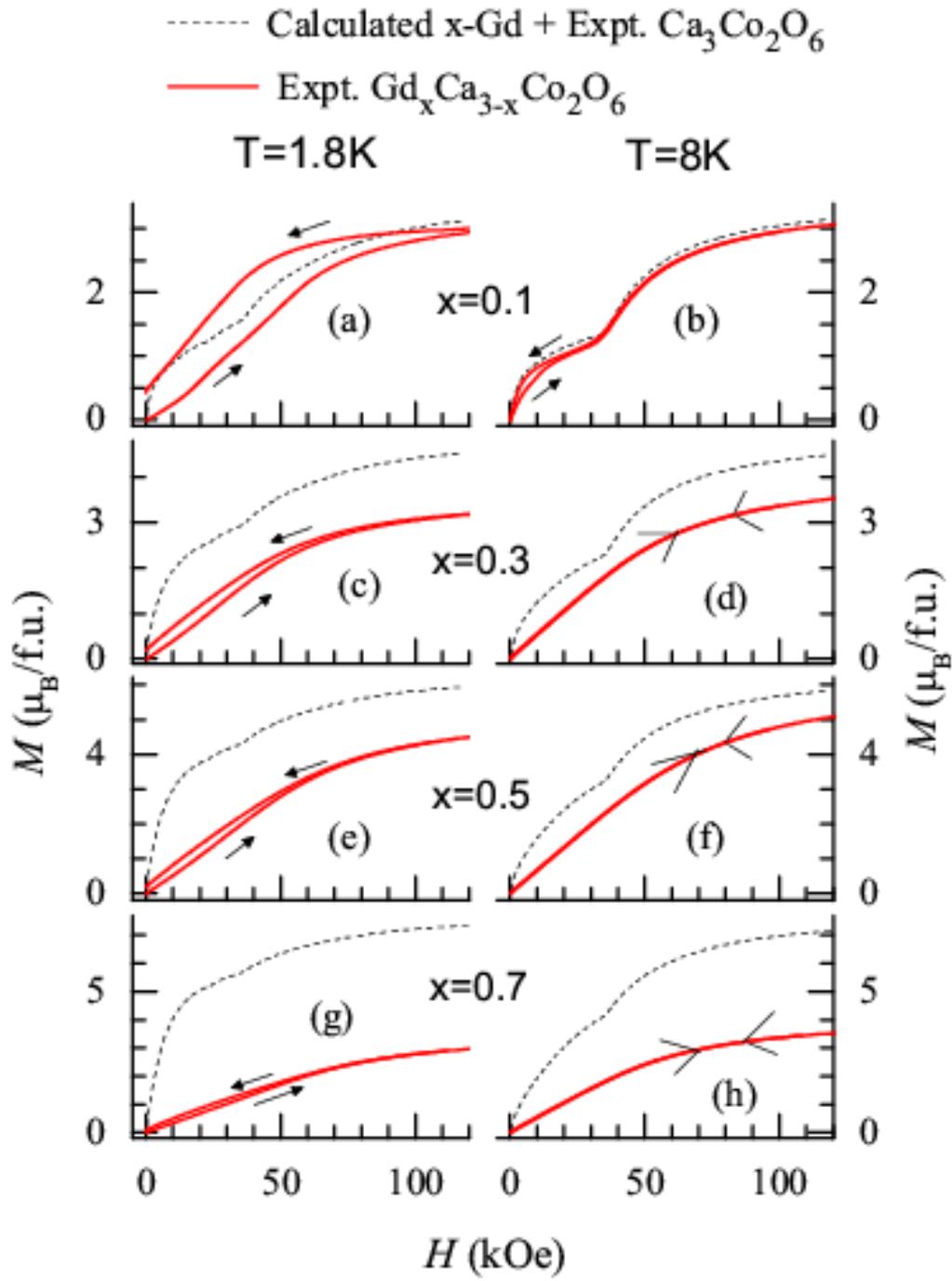

Figure 4

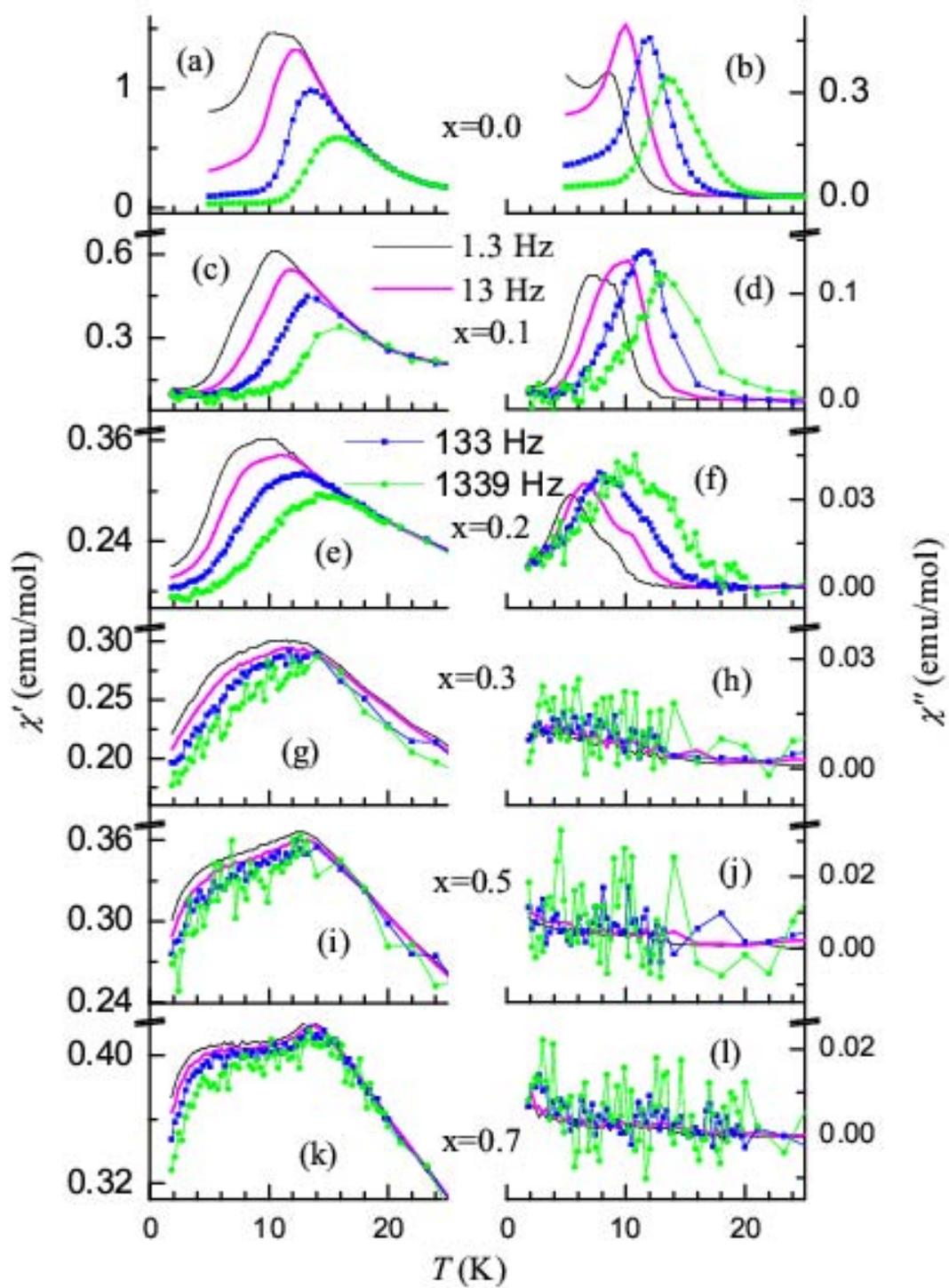

Figure 5

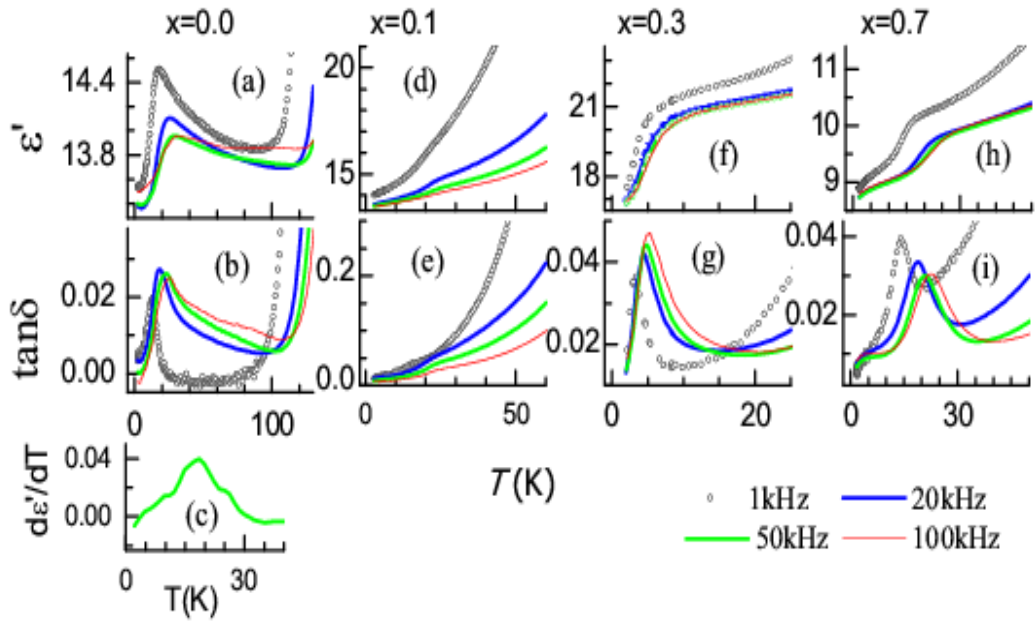

Figure 6

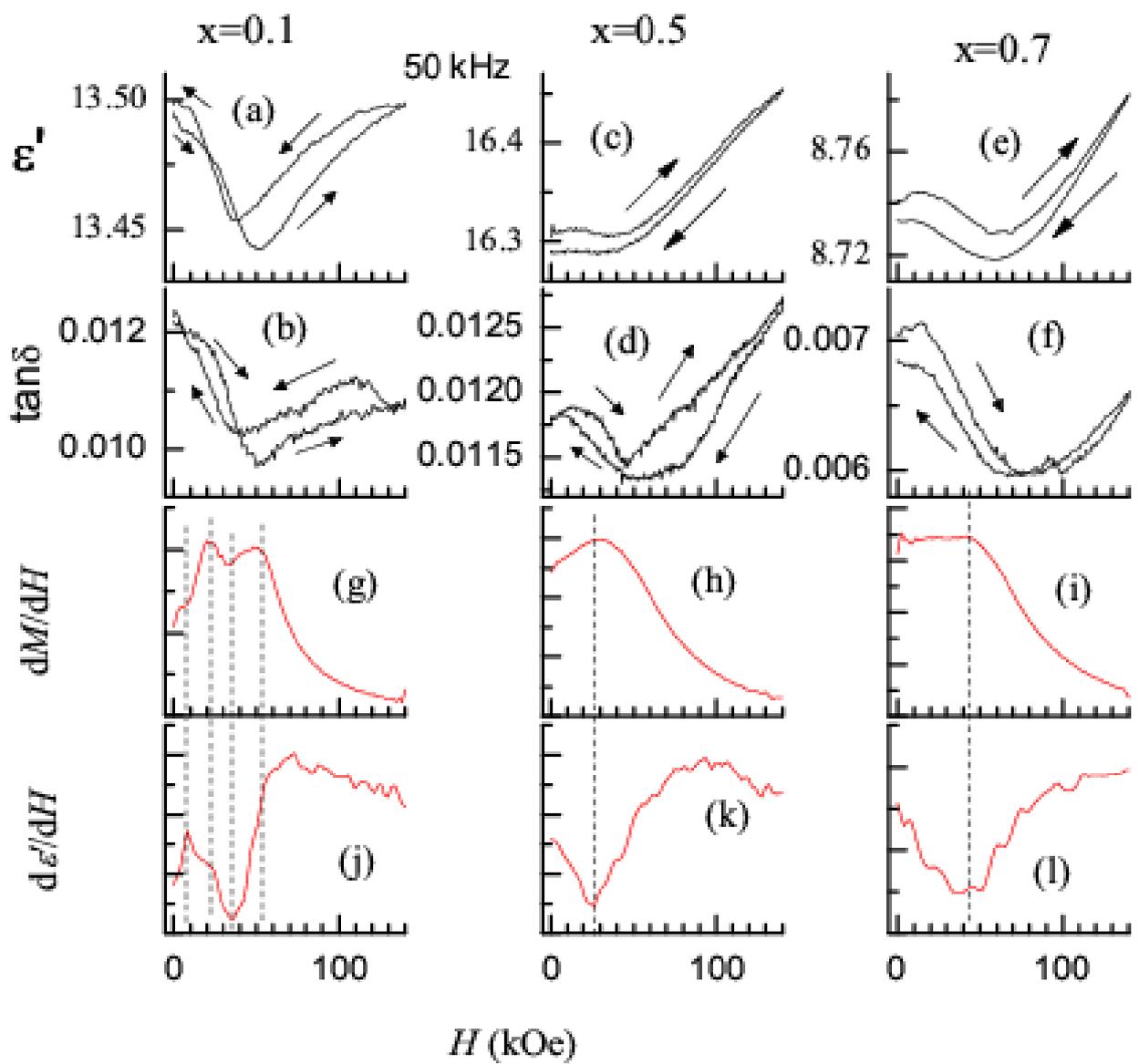

Figure 7